\documentclass[12pt]{article}
\title{AGK in the parton model}
\author{Alexey V. Popov\thanks{email: avp@novgorod.net} \\ \\ 
\small{\emph{Novgorod State University, B. S.-Peterburgskaya Street 41,}}\\
\small{\emph{Novgorod the Great, Russia, 173003}}
} 
\date{}

\begin{document}
\maketitle
\abstract{
We propose pure quantum parton model of soft interactions. The model is based on soft partons which interact eikonally and
locally in a transverse plane. 
We directly construct S-matrix operator and partons Fock space using developed method of discretization of a transverse plane.
Counting inelastic states we explicitly derive AGK cancellations and cutting rules from relation 
between inclusive and exclusive parton distributions. We also discuss black disk behavior from the point of view of our model. 
}
\section{Introduction}
AGK cutting rules \cite{AGK} are very important part of high energy phenomenology. Usually it formulated using
reggeon diagrams.  The main problem is the lack of a theory which can say how exactly study multipomeron vertexes. 
Most conventional and naive approaches are eikonal and quasi-eikonal models. But there is no theoretical motivation to use this models on
fundamental level. More predictive and intuitive model can be obtained using partons. We can associate multipomeron vertex with
existence of multiparton states in a hardron wave function. Key task here is to find and define parton properties. Then to build the theory using
parton as elementary building block.

There are numerous works about regimes with very high energies and high partons density. 
Formation of a black disk causes especial interest. It has been noticed \cite {Kancheli} that a t-channel cut of arbitrary 
reggeon  diagram gives only pure n-reggeon exchange and we can absorb upper and lower diagram parts to the projectile and target 
respectively. This picture can be naturally reformulated in terms of partons. Hardron is compound object builded from partons. 
If we boost hardron to high $Y$ then in its wave function there will be new partons which is strongly correlated with primary partons. 
During the collision, partons evolution is frozen and we deal only with given wave function represented as certain vector in Hilbert space.
The main observation is that the evolution and scattering is distinct and independent parts of the theory. This this similar to the high
energy QCD where scattering is Wilson lines and evolution is a Hamiltonian-driven task.
Obviously, the theory of elemental soft partons interactions should be developed. In this paper we propose the approach to such model
and explicitly derive AGK cutting rules.

\newcommand{\ket}[1]{\ensuremath{|#1\rangle}}
\newcommand{\bra}[1]{\ensuremath{\langle#1|}}
\newcommand{\braket}[2]{\ensuremath{\langle#1|#2\rangle}}
\section{The model}
Our model based on the old Good and Walker idea \cite{Good_Walker}. Consider interaction of the hardron projectile 
and hardron target at high energy. The projectile, eigenstate of energy, being composite, can be described
as superposition of states which are eigenstates of quasielastic scattering. After scattering, projectile state 
is not simply proportional to the initial projectile. It is superposition of different observable particles. This gives a possibility
of inelastic diffraction process. In other words, in scattering process coherence of projectile lost and disbalance of wave function 
observed as possibility of decay of the projectile. 

Second ingredient of our model is a partons. In soft hardrons scattering phenomenology a colorless parton models are widely applied. 
For example, the pomeron is qualitative described by ladder-like diagram. It is natural to conjecture that the eigenstates of scattering which 
leads to diffraction and partons are closely related. Our hypothesis -- that they are the same. More precisely, each parton in a projectile
wave function is eigenstates of quasielastic scattering. Parton can scatter elastically or pure absorptive forming something like multiperipheral 
state where produced particles populate midrapidity region. The Fock space of parton states is equivalent to the space of all eigenstates of scattering.

Third ingredient of our model is a evolution. Instead of studying usual limit $Y\rightarrow\infty$ we study relatively small $Y\sim 1$ 
where loop corrections can be neglected. If we increase~$Y$ then the projectile wave function will change.
This happens due to parton emission into new opened region in phase space. So instead of increasing $Y$ we can simple change projectile
wave function and replace old projectile by this new state. The situation is very similar to the studying of the high energy
scattering in the perturbative QCD where JIMWLK evolution equation can be derived by boosting projectile \cite{Kovner_05}.

Fourth ingredient of our model is the eikonal approximation for multiparton scattering. This clause is also QCD-motivated. In QCD amplitude of scattering at 
moderate $Y$ can be computed by evaluating Wilson lines of projectile and target partons: $\exp(ig\int A_\mu(x) dx^{\mu})$, where $A_\mu$ is 
the classical gauge field generated by all partons \cite{Balitsky_01}. It is important that the energetic partons moves along straight lines. 
From general quantum field theory framework this picture can be explained as quasiclassical approximation. Scattering amplitude can be
written as functional integral $\int e^{iS[\varphi]}D\varphi$, where $\varphi(x)$ - some fields related with relevant degrees of freedom.
$\varphi$ must obey appropriate border conditions which correspond to initial and final energetic partons of projectile and target.   
Quasiclassical solution $\varphi_0(x)$ is a field configuration with extremal action which obey all necessary boundary conditions. 
Scattering amplitude can be computed as $e^{iS[\varphi_0]}$.

Fifth ingredient of our model is a transverse locality. It means that classical field of each soft parton fast decrease with transverse
distance, in opposite to QCD where fields decrease polynomially. If the hardron is dilute then the fields of parton in not
overlap with each other. Geometrically, this means that parton size much smaller than average distances
between partons and partons can be treated as point-like particles in a transverse plane.  
Hardron transverse extension totally associated with wave function of constituent partons, not with parton 
fields. This is similar to hydrogen atom where point-like electron have broad wave function.
It should be stressed that high parton density is not automatically equivalent to high parton number where our model is not
applicable. For example, if parton localized in small transverse area then locally density is high but there is only one parton.

Projectile state $\ket{\Psi_P}$ can be written as vector in the Fock space generated by action of parton creation operator $a^\dag_1(x)$ on the vacuum, 
where $x$ - transverse position. Respectively target state $\ket{\Psi_T}$ generated by action of $a^\dag_2(x)$.

Main building block in our model is elementary two parton $S$-matrix. 
Consider collision of two partons with positions overlapped in transverse plane.
Initial state is 
\begin{equation}
\ket{\Psi_i}=a^\dag_1 a^\dag_2 \ket{0}
\end{equation}
Due to all five our assumptions we write
\begin{equation} \label{eq_10}
S\ket{\Psi_i}=\mu \ket{\Psi_i} + \nu c^\dag\ket{0}
\end{equation}
where $\mu$,$\nu$ is some constants, and $c^\dag$ is operator of creation of inelastic state which we will call as c-modes.
We assume that the created inelastic state is also transverse localized and we can create its with local operators $c^\dag(x)$.
We can motivate transverse localization of c-modes by following reasons. From quasiclassical approximation point of view the
c-mode is just another quasiclassical solution type of the field equations with special border conditions. There are no reasons 
to expect a discontinuous transverse jump of a classical solution. Another argument comes from color flux tube phenomenology.
Soft partons receives a color charge in collision and color flux tube between patrons is formed. Such tube is also should be
transverse localized.

Constants $\mu$ and $\nu$ must obey unitary relation
\begin{equation}
|\mu|^2+|\nu|^2=1
\end{equation}  
The next step is to find secondary quantized version of equation (\ref{eq_10}) where arbitrary number of partons can present in the initial state.

\section{Discretization of a transverse plane}
 Partons Fock space can be constructed by creation operators obeying 
\begin{equation}
[a(x),a^\dag (y)]=\delta(x-y)
\end{equation}
Although we use boson statistics for partons, our subsequent calculations is not depend on parton statistic type.
Each parton can have arbitrary transverse position which is continuous variable. So this space is not separable. 
If we deal only with finite partons number then the probability to have two partons in some point has zero measure. 
So the occupation numbers can have value only 0 and 1. Note that this is similar to studying fermions.

In order to do more rigorous and simple math we approximate original Fock space by Fock space constructed
on discrete transverse plane where partons can sits only on lattice nodes. Then we can tend lattice spacing to zero.
Original parton position we replace by nearest node position. 
In our model if we change lattice size then the the number of partons is not change.
But in the QCD this is never true due existence of RG group flow and DGLAP evolution.
Moreover, QCD parton Fock space can be correctly formulated only on lattice because if we tend lattice spacing to zero then
the parton density grow to infinity. Lattice spacing play role of momentum cutoff. Changing cutoff we go from a coarse-grained description
to a fine-grained description.
We introduce discrete parton creation operators obeying
\begin{equation}
[a_i,a^\dag_j]=\delta_{ij}
\end{equation} 
where $i,j$ - indexes of lattice nodes.
For arbitrary operator $\hat O$ we denote as $:\hat O:$ normal ordered operator, as $\hat O_i$  -- its discrete version, as $\hat O(x)$ -- 
its continuous versions.
General correspondence rule between continuous and discrete versions of operators is  
\begin{equation} \label{eq_11}
a_i= a(x_i) \sqrt {\Delta s}
\end{equation}
\begin{equation}
N_i=N(x_i) \Delta s
\end{equation}
where $\Delta s$ - area associated with lattice node and $N_i=a^\dag_ia_i$ operator of parton number.
It is should be stressed that the (\ref{eq_11}) has meaning only as approximation 
of matrix elements of operators by matrix elements of its discrete version. It is straightforward to show that
\begin{equation}
\sum_i N_i=\int N(x) dx
\end{equation}
In our model if the lattice spacing is sufficiently small then in each node can be no more than one parton.
In the continuum limit the probability to have at least two patrons with equal coordinates is neglible small.
If maximum occupation number is one then we have 
\begin{equation} \label{eq_9}
:N_i^2:=0
\end{equation}
This is very important property for our subsequent calculations.
As example we construct projector on the Fock vacuum. 
\begin{equation} 
P_0=\prod_i(1-N_i)
\end{equation}
This operator give zero if at least one parton exist. Next we want to find continuous version of this operator.
With help of (\ref{eq_9}) we can write
\begin{equation} \label{eq_17}
1-N_i=:e^{-N_i}:
\end{equation}
\begin{equation}\label{eq_16}
P_0=:e^{-\sum N_i}:=:e^{-\int N(x) dx}:
\end{equation}
Now we can construct secondary quantized $S$-matrix. Formula (\ref{eq_10}) can be rewritten as 
\begin{equation} \label{eq_12}
S_i=1-(1-\mu)N_{1,i}N_{2,i}+\nu c_i^\dag a_{1,i} a_{2,i}
\end{equation}
Equation (\ref{eq_12}) is very natural and valid only if the lattice spacing greater than parton size.
Continuous version of full $S$-matrix 
can be derived applying derivation method of (\ref{eq_16}) to (\ref{eq_12}) 
\begin{equation} \label{eq_14}
S=\prod_i S_i=e^{-(1-\mu)\Delta s\int N_1(x) N_2(x) dx+\nu \sqrt{\Delta s} \int a_1(x) a_2(x) c^\dag(x) dx}
\end{equation}
When tending $\Delta s$ to zero to get continuous version we have a problem that $\Delta s$ can be smaller than the parton size $s_0$.
Recall that for each $\Delta s$ we only have approximation for matrix elements and smaller $\Delta s$ gives better approximation.
So we take $\Delta s=s_0$ as best available approximation. If we go to $\Delta s<s_0$ then locality in (\ref{eq_12}) will lost.
When we calculate amplitude of realistic process with given momentum transfer we effectively average parton densities over 
distances of scale of inverse momentum. So it is not necessary to have very fine-grained description. Finally, we have
\begin{equation} \label{eq_20}
S=e^{-A\int N(x) dx+\lambda \int a(x) c^\dag(x) dx}
\end{equation}
where $A=(1-\mu)s_0$, $\lambda=\nu \sqrt{s_0}$ and $a(x)=a_1(x)a_2(x)$, $N(x)=N_1(x)N_2(x)$

\section{Cutting rules}
Now our task is to explore how many c-modes exist in the final state
\begin{equation}
 \ket{\Psi_f}=S\ket{\Psi_i}
\end{equation}
We define elastic scattering amplitude as
\begin{equation}
M=\bra{\Psi_i} 1-S \ket{\Psi_i}
\end{equation}
\begin{equation} \label{eq_18}
M^{(n)}=(-1)^{n+1} A^n f^{(n)}
\end{equation}
where we defined inclusive partons spectrum in the initial state 
\begin{equation}
f^{(k)}(x_1\ldots x_k)=\bra{\Psi_i} N(x_1) \ldots N(x_k)\ket{\Psi_i}=f^{(k)}_1(x_1\ldots x_k)f^{(k)}_2(x_1\ldots x_k)
\end{equation}
\begin{equation} \label{eq_6}
f^{(n)}=\frac{1}{n!}\int f^{(n)} (x_1 \ldots x_n)  dx_1 \ldots dx_n
\end{equation}
and where $f_1$ and $f_2$ is projectile and target parton inclusive distributions respectively.
Note that the factor $(-1)^{n+1}$ lead to so called shadowing corrections.
Definition of the total cross section is
\begin{equation}
\sigma_{tot}=\sum_n \left|\bra{n} 1-S \ket{\Psi_i}\right|^2
\end{equation}
Using unitary $S^\dag S=1$ and complete basis $\ket{n}$ we can obtain optical theorem
\begin{equation} \label{eq_19}
\sigma_{tot}=2ReM
\end{equation}
Inclusive spectrum of the c-modes in the final state is 
\begin{equation}
f_c^{(k)}(x_1\ldots x_k)=\bra{\Psi_f} N_c(x_1) \ldots N_c(x_k)\ket{\Psi_f}
\end{equation}
where $N_c(x)=c^\dag(x)c(x)$.
We define generating functional for inclusive spectrum
\begin{equation} 
F_c[u]=\sum_{n=0}^\infty \frac{1}{n!}\int f_c^{(n)}(x_1 \ldots x_n) u(x_1) \ldots u(x_n) dx_1 \ldots dx_n
\end{equation}
This functional can be equivalently rewritten as 
\begin{equation} \label{eq_2}
F_c[u]=\bra{\Psi_f} :e^{\int u(x) N_c(x) dx}: \ket{\Psi_f}=
       \bra{\Psi_i}\prod_i S^\dag_i \left(1+u_i N_{c,i}\right) S_i\ket{\Psi_i}
\end{equation}
Using (\ref{eq_12}) and relation (\ref{eq_17}) we obtain
\begin{equation}
F_c[u]=\bra{\Psi_i}\prod_i \left(1+|\nu|^2 u_i N_{1,i}N_{2,i} \right) \ket{\Psi_i}=\bra{\Psi_i} :e^{\int |\lambda|^2 u(x) N(x) dx}: \ket{\Psi_i}
\end{equation}
and therefore
\begin{equation} \label{eq_13}
f^{(k)}_c(x_1\ldots x_k)=|\lambda|^{2k} f^{(k)}(x_1\ldots x_k)
\end{equation}
Equations (\ref{eq_13}) is the direct analog of the well-known AGK cancellations.
From (\ref{eq_18}) we see that the $f^{(k)}$ gives the $k$-fold parton elastic scattering. 
Equation (\ref{eq_13}) says that $k$-fold inclusive inelastic spectrum is solely related 
to $k$-fold contribution to the elastic amplitude.

Now we want to find exclusive distribution of c-modes. 
In the our model exclusive spectrum gives multiplicity distribution in the final state.
It can be explicitly constructed as
\begin{equation} \label{eq_1}
g^c_{i_1\dots i_n}=\bra{\Psi_f}   N_{c,i_1} \ldots N_{c,i_n}  \prod_{i\neq i_k} (1-N_{c,i})  \ket{\Psi_f}
\end{equation}
Operator $1-N_{c,i}$ plays role of the vacuum projector at given point $x$. So we select only states with precisely $n$ c-modes 
placed in nodes $i_1 \ldots i_n$.
Using definition of inclusive distribution (\ref{eq_2}), equation (\ref{eq_1}) in its continuous version can be expressed as 
\begin{equation} \label{eq_4}
g^c_n(x_1 \ldots x_n)=\left. \frac{\delta^n F_c[u]}{\delta u(x_1) \ldots \delta u(x_n)} \right|_{u=-1}
\end{equation}
Hence the generating functional $G_c[u]$ for exclusive distribution can be calculated using inclusive distribution
\begin{equation}
G_c[u]=F_c[u-1]
\end{equation}
If we define
\begin{equation}
g_n=\frac{1}{n!}\int g^{(n)}(x_1\ldots x_n) dx_1 \ldots dx_n
\end{equation}
then the normalisation is
\begin{equation}
\sum_{n=0}^\infty g_n=G_c[1]=1
\end{equation}
Probability to have $n$ c-modes in the final state equals to $g^c_n$. 
From (\ref{eq_4}) we can directly calculate integrated exclusive distribution for c-modes
\begin{equation} \label{eq_5}
g^c_n=\sum_{k=n}^\infty g_n^{(k)}
\end{equation}
\begin{equation}
g_n^{(k)=(-1)^{k-n} \frac{n!}{(k-n)! k!} f_c^{(k)}}
\end{equation}
where we defined $g_n^{(k)}$ as the contribution to $g^c_n$ from $f_c^{(k)}$.
Now it is useful to relate probabilities $g_n$ to cross sections of subprocesses.
Let $\sigma_k^{(n)}$ is the contribution to the cross section of create $k$ number of c-modes from $n$ order.
If $k>0$ then we obviously have
\begin{equation}
\sigma_n^{(k)}=g_n^{(k)}
\end{equation}
and for $n=0$ we use the following identity 
\begin{equation}
\sigma_{tot}^{(k)}=\sigma_0^{(k)}+\sigma_1^{(k)}+\ldots\sigma_k^{(k)}=\sigma_0^{(k)}-g_0^{(k)}+\sum_{n=0}^k g_n^{(k)}
\end{equation}
From (\ref{eq_5}) it is easy to check that the sum in last expression is zero and we obtain 
\begin{equation}
\sigma_0^{(k)}=g_0^{(k)}+\sigma^{(k)}_{tot}=g_0^{(k)}+2Re M^{(k)}
\end{equation}
where we used optical theorem (\ref{eq_19}).
Substituting $f_c^{(k)}$ from (\ref{eq_13}) into (\ref{eq_5}) we finally have
\begin{equation} \label{eq_3}
\begin{array}{l}
\sigma_n^{(k)}=(-1)^{n+1} C_k^n  D^{(k)} \\
\sigma_0^{(k)}= 2Re M^{(k)} - D^{(k)}
\end{array}
\end{equation}
where
\begin{equation} \label{eq_7}
D^{(k)}=(-1)^{k+1} |\lambda|^{2k}f^{(k)}
\end{equation}
Equations (\ref{eq_3}) is well known AGK cutting rules (integrated) \cite{AGK}.
We can easy check the following property  
\begin{equation}
\sum_{n=0}^k \sigma_n^{(k)} = \sigma^{(k)}_{tot}
\end{equation}
From equations (\ref{eq_3}) we can derive Pumplin bound for cross section of the diffractive dissociation \cite{Pumplin}. 
By definition 
\begin{equation}
\sigma_0=\sum_{k=1}^\infty\sigma_0^{(k)}=\sigma_{el}+\sigma_{diff}
\end{equation}
from (\ref{eq_3}) we have
\begin{equation} \label{eq_8}
\sigma_0=\sigma_{tot}-\sum_{k=1}^\infty f^{(k)}(-1)^{k+1}|\lambda|^{2k}
\end{equation}
Pumplin bound valid only if $|A|\ll 1$ and $Im A\ll Re A$, where $A$ has been defined in (\ref{eq_20})
In this limit unitary relation have form
\begin{equation}
2Re A=|\lambda|^2
\end{equation}
\begin{equation}
|\lambda|^{2k}=2^{k-1}2Re A^k=\frac{1}{2}2Re (2A)^k
\end{equation}
Substituting this into (\ref{eq_8}), we obtain
\begin{equation}
\sigma_0=\sigma_{tot}-\frac{1}{2} 2Re \bra{\Psi_i} 1- e^{-\int2AN(x)dx} \ket{\Psi_i} \leq \frac{1}{2}\sigma_{tot}
\end{equation}

\section{Black disk states}
Note that the eikonal model is equivalent to factorization of inclusive partons distributions
\begin{equation} \label{eq_15}
f^{(k)}(x_1 \ldots x_k)=f^{(1)}(x_1) \ldots f^{(1)}(x_k)  
\end{equation}
From exclusive distribution point of view this relation can be viewed as possible existence of arbitrary number of partons in the wave 
function. And it is well known that in the final state multiplicity have Poisson-like distribution. This situation usually signals about
some sort of chaos and can arise if statistically many degree of freedom involved in process. We only can hope that this scenario will 
arise in limit of very high $Y$. But in case of a dilute hardron there is no reasons to expect such factorization. 
Moreover, in case of finite number of partons there are no multipomeron exchanges of order higher than partons number.

Consider black disk limit. At first we must clarify meaning of the term "black disk". Usually it defined as target with vanishing 
elastic S-matrix for arbitrary projectile. This is rather unpleasant requirement because it is difficult to require exactly $S=0$ by 
physical reasons. Instead we use more weak version and define black disk as target hardron satisfying to relations (\ref{eq_15}) with
some accuracy. This definition can be used in dilute regime. 
If projectile is the black disk too then from (\ref{eq_14}) elastic S-matrix has form 
\begin{equation}
S_{bd}=e^{-A\int f_1(x)f_2(x)dx}
\end{equation}
More precisely speaking, our definition of black disk should be called as gray disk. True black disk arises at high partons densities.

Now we show some statistical arguments for inevitability of formation of black disk states. At first we consider $N$ partons uniformly
distributed in area $S$ in the transverse plane. This state can be correctly described by density matrix $\rho$. 
If state have exactly $N$ partons in the area $S$  then
\begin{equation}
\rho \ket{x_1 \ldots x_n}=\frac{1}{Z}  \ket{x_1 \ldots x_n}
\end{equation}
and it is equal to zero in other cases. Normalization constant is $1/Z=N!/S^N$. Then we have 
\begin{equation}
f^{(1)}(x_1)=  \langle \hat N(x_1) \rangle = Sp [\hat N(x_1) \rho]=\frac{N}{S}
\end{equation}
\begin{equation}
f^{(2)}(x_1,x_2)=\frac{N^2}{S^2}\left(1-\frac{1}{N}\right)
\end{equation}
\begin{equation}
f^{(3)}(x_1,x_2,x_3)=\frac{N^3}{S^3}\left(1+\frac{3}{N}-\frac{3}{N^2}\right)
\end{equation}
and so on. We see that if $N$ is large
than lowest inclusive distributions obey factorizations (\ref{eq_15}). So considered state is good approximation to 
a black disk. Such states can describe nucleus wave function in term of partons, at least locally.

Next argument comes from entropy consideration. Consider large hardron state $\ket{\Psi}$.
Let take small area $S_1$ within hardron and consider subsystem in $S_1$ as quantum system.
Hilbert space can be factorized as tensor product $H'\otimes H_1$, where $H_1$ - Hilbert space 
for subsystem $S_1$ and $H'$ for subsystem $S'=S-S_1$. If we want to calculate averages of operators acted in $H_1$ then it is
useful to introduce density matrix $\rho_1$ for subsystem $S_1$.
\begin{equation}
\rho_1= Sp'  \ket{\Psi} \bra{\Psi}
\end{equation}
where trace taken over $H'$ space. Having non trivial density matrix we can evaluate entropy 
\begin{equation}
E_1=-Sp_1(\rho \ln\rho)
\end{equation} 
where trace taken over $H_1$ space. Now we use general property of chaotic systems that entropy tends to be additive.
We assume that the hardron is locally chaotic. Equivalently, $E_1$ proportional to $S_1$.
Such reasoning similar to derivation of Gibbs distribution from microcanonical ensemble. If the entropy of two small subsystems
$S_1$ and $S_2$ is additive
then the density matrix of the subsystem $S_{1+2}$ is a the product $\rho_{1+2}=\rho_1 \otimes \rho_2$. 
In calculations of $f^{(2)}(x_1,x_2)$ we can take two subsystems $S_1$ and $S_2$ where $x_1\in S_1$ and $x_2\in S_2$.
\begin{equation}
f^{(2)}(x_1,x_2)=\langle N(x_1) N(x_2) \rangle= Sp_1(N(x_1)\rho_1) Sp_2(N(x_2)\rho_2)=\langle N(x_1) \rangle \langle N(x_2) \rangle
\end{equation}
So we have factorization (\ref{eq_15}), not for whole hardron but for small areas within it. This can be applicable for
collision of small dilute projectile with large target.

\section{Discussion}
In this paper we have derived AGK cutting rules in the framework of constructed parton model. Remarkable that the result comes 
from pure algebraic relation between inclusive and exclusive distributions for inelastic states.  There are some unclarified
questions here. 

First question is how inelastic c-modes converts into realistic hardrons. At first sight in spirit of color flux tube and 
multiperipheral states we can view this state as state where partons populates whole available rapidity interval but 
only in same transverse position.  Then using realistic hardrons wave functions we must project final state on observable particles.
So partons from different c-modes can compose into one hardron. 

Second question is about nature of our partons. How relate partons to constituent quarks or to string modes in 
dual resonance model? It is not clear. But from many examples in field theories we see that excited field modes near some vacuum have
Fock space form. Crucial assumption is only transverse locality in a dilute regime.

Next question is about evolution equation for hardron wave function. In spirit of a QCD calculations at first approximation we can 
write result of infinitesimal boost as
\begin{equation} \label{eq_21}
\ket{\Psi'}=\left (
\begin{array}{l}
1+\int G(z-x) a^\dag(z) N(x) dx dz +\\
\phantom{1}+ \int G^*(z-x)G(z-y) N(x) N(y) dx dy dz
\end{array}
 \right)\ket{\Psi}
\end{equation}
where universal function $G(x)$ is a patron emission amplitude. It is directly related to the pomeron intercept and slope. 
This function must be extracted from more fundamental theory.
Equation (\ref{eq_21}) is the direct analog of the JIMWLK equation in the QCD.
The pomeron here is just asymptotic of the scattering amplitude. 

In pomeron calculations widely used eikonal model with gauss vertexes
\begin{equation}
N(k_1\ldots k_n)=g^ne^{-R^2\sum k_i^2}
\end{equation}
It can be shown that in the parton model this equivalent to gauss distributions too
\begin{equation}
f^{(1)}(x)\sim e^{-ax^2}
\end{equation}
Unclear question here is about partonic interpretation of reggeon signature factor which is not depend on $Y$ and on hardron
properties.
\section*{Acknowledgments}
We thank N.V. Prikhod'ko for feedback and useful remarks.

\end{document}